\documentclass{article}
\usepackage{spconf,amsmath,graphicx,hyperref}
\usepackage{textcase}
\usepackage{makecell}

\usepackage{xcolor}

\title{Hallucination Reduction for LLM-based Audio Understanding via Multimodal Direct Preference Optimization}

\name{Bebe Cosgrove$^{\star}$\thanks{This research was completed during Bebe Cosgrove's NSF-funded Research Experiences for Undergraduates (REU) internship (Award \#2349121: Computing for Health and Well-being) at the University of Iowa.}
\qquad Aaron Isidore Grace$^{\dagger}$ \qquad Weiran Wang$^{\dagger}$}
  
\address{$^{\star}$Smith College, Northampton, USA \\
  $^{\dagger}$University of Iowa, Iowa City, USA \\
  \texttt{bcosgrove@smith.edu} \quad \texttt{aaron.grace@uwaterloo.ca} \quad \texttt{weiran-wang@uiowa.edu}
  }

\begin{document}
\ninept
\maketitle
\begin{abstract}
Large Audio Language Models (LALMs) are prone to hallucinating and over-relying on text priors when simultaneously presented with audio and text inputs. To mitigate these hallucinations, we propose utilizing the multimodal Direct Preference Optimization (mDPO) objective, which forces the model to ground its generation in the acoustic input by contrasting intact and distorted audio counterparts. We extend this preference learning framework to the audio domain by applying a variety of acoustic perturbations. Evaluating the Qwen2-Audio backbone across the DCASE 2025 Challenge and AH Existence datasets, we demonstrate that extending mDPO to LALMs significantly enhances temporal reasoning in the complex DCASE dataset, and improves performance on basic existence verification in the AH benchmark. We identify temporal reversal, frequency masking, and random noise as the most effective perturbations. Ultimately, our approach achieves an absolute accuracy improvement of 14.0\% on the DCASE 2025 dataset and 27.4\% on AH Existence.
\end{abstract}
\begin{keywords}
large audio-language models, audio hallucination, direct preference optimization
\end{keywords}
\section{Introduction}
\label{sec:intro} 
Large Audio-Language Models (LALMs)~\cite{deshmukh2023pengi, tang2024salmonn,  gong2024listen, hu2024wavllm, lu2025desta25Audio} allow for the capabilities of language models to extend to scenarios where audio understanding is necessary. Despite their growing popularity and importance, LALMs have routinely been hallucinating when tested on unseen audio data~\cite{wang-etal-2025-audio, kuan2025can, kuan2024understanding, zhao2026halluaudio, cheng2025ahabench, sungbin2025avhbench}. This vulnerability largely stems from the fact that foundational LLMs are predominantly trained on textual data; consequently, when presented with simultaneous audio and text, LALMs often exhibit a strong bias toward the text modality, ignoring acoustic cues and thereby generating hallucinations. 

Several training-free approaches have been proposed to mitigate audio hallucinations, such as prompt engineering~\cite{hsu2025reducing, grace2026adaptive}, chain-of-thought reasoning~\cite{kuan2025can}, and contrastive decoding~\cite{hsu2025reducing, grace2026adaptive}. Contrastive decoding, for example, reduces hallucinations at inference time by contrasting standard predictions with a perturbed audio branch. However, these training-free methods are fundamentally limited; they rely on heuristic adjustments during generation and do not natively correct the model's underlying multimodal representations or linguistic biases.

To address this over-reliance on text priors, we draw inspiration from the vision-language domain, where Conditional Preference Optimization for Multimodal Large Language Models (mDPO)~\cite{wang-etal-2024-mdpo} was developed to encourage models to attend to the image input. mDPO was originally proposed as an extension of Direct Preference Optimization (DPO)~\cite{rafailov2023direct} to multimodal models. 
DPO focuses on tuning the model to mimic human preferences while bypassing the need to train a separate reward model and avoiding the computational overhead of reinforcement learning algorithms like PPO~\cite{bai2022constitutional, christiano2017deep}. 
Utilizing the methodology of DPO, mDPO specifically encourages the model to look at the visual input by contrasting a regular image with a slightly distorted version of the same image, both paired with identical text. Because the text inputs are identical, the model is forced to rely on the image to distinguish between the chosen and rejected responses~\cite{wang-etal-2024-mdpo}.

In our work, we propose extending mDPO to the audio domain. Though the original mDPO uses visual perturbations such as cropping, these cannot be directly applied to audio. Therefore, to extend mDPO to audio LLMs, we take inspiration from recent work on contrastive decoding~\cite{hsu2025reducing} and explore which audio perturbations are most effective at eliminating hallucinations. In our experiments, we evaluate several audio perturbations, including random Gaussian noise, complete audio masking, temporal reversal, as well as time and frequency masking. We conduct our experiments utilizing the Qwen2-Audio7B-Instruct~\cite{yang2024qwen2technicalreport} architecture on two prominent benchmarks: Task 5 of the DCASE (Detection and Classification of Acoustic Scenes and Events) 2025 Challenge and the Audio Hallucination (AH) benchmark. On the DCASE 2025 dataset, a combination of perturbations achieved a 14.0\% absolute improvement; similarly, on the AH Existence subset, the reverse perturbation yielded an absolute accuracy improvement of 27.4\%.

\begin{figure*}[!t]
    \centering
    \includegraphics[width=0.8\linewidth]{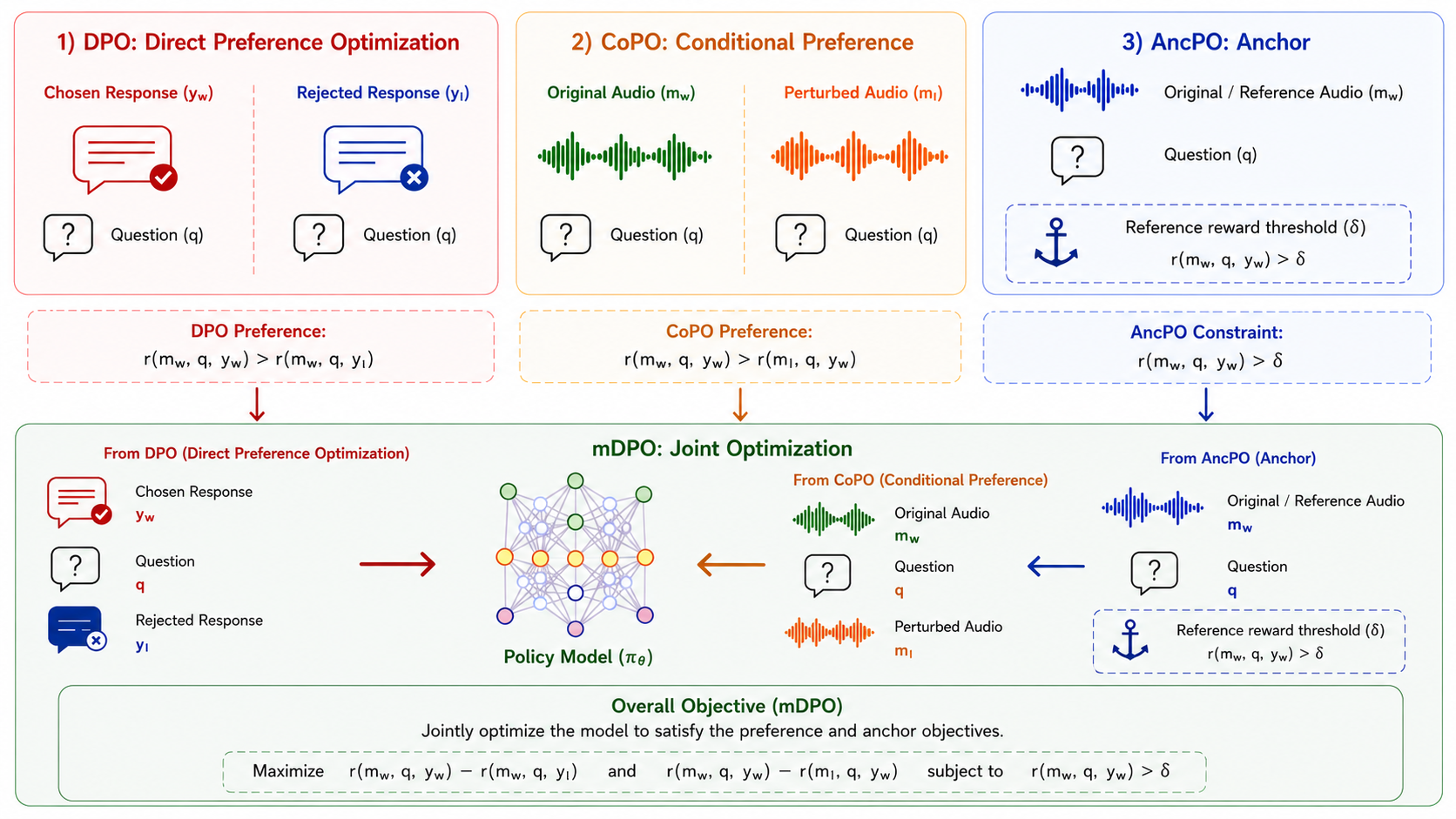}
    \caption{Overview of the mDPO framework for audio hallucination reduction.}
    \label{fig:mdpo_audio}
\end{figure*}

\section{Methods}
\label{sec:methods}
\subsection{Audio-Grounded mDPO Objective}
This section details the formulation of the mDPO objective function, an overview of which is visually summarized in Figure~\ref{fig:mdpo_audio}. Let $q$ denote a textual query, $y_w$ the preferred response, and $y_l$ the rejected response. Furthermore, let $m$ represent the multimodal input, where $m_w$ denotes the original, unperturbed input and $m_l$ denotes its distorted counterpart. The comprehensive mDPO loss is composed of three distinct terms, defined as follows.
\\
\subsubsection{Standard DPO Term}
The foundational component of the objective is the standard DPO loss, $\mathcal{L}_{\mathrm{DPO}_m}$, which aligns the model's generation with human preferences and ensures that baseline instruction-following capabilities are preserved during multimodal fine-tuning~\cite{rafailov2023direct}. Given a parameterized policy $\pi_{\theta}$ and a frozen reference policy $\pi_{\mathrm{ref}}$, this term optimizes the network by maximizing the implicit reward margin between the preferred response $y_w$ and the rejected response $y_l$~\cite{wang-etal-2024-mdpo}. 
\begin{equation*}
\begin{aligned}
\mathcal{L}_{\mathrm{DPO}}
= -\log \sigma \Bigg(
&\beta \log
\frac{\pi_{\theta}(y_w \mid m,q)}
     {\pi_{\mathrm{ref}}(y_w \mid m,q)}
-
\beta \log
\frac{\pi_{\theta}(y_l \mid m,q)}
     {\pi_{\mathrm{ref}}(y_l \mid m,q)}
\Bigg)
\end{aligned}
\end{equation*}
Here, $\sigma$ denotes the logistic sigmoid function, mapping the reward difference into a probability space, and $\beta>0$ is a hyperparameter. 

\subsubsection{Conditional Preference Optimization Term}
To explicitly penalize multimodal hallucination, the Conditional Preference Optimization term, $\mathcal{L}_{\mathrm{CoPO}}$, forces the model to ground its generation in the provided multimodal context rather than relying on language priors~\cite{wang-etal-2024-mdpo}. Given identical textual queries $q$ and preferred responses $y_w$, this term maximizes the implicit reward margin between the intact multimodal input $m_w$ and its perturbed counterpart $m_l$. Because the textual conditions are identical, the model must actively attend to the multimodal input to successfully differentiate the two scenarios. While the original mDPO framework utilized visual distortions such as image cropping, we adapt this objective to the audio domain by defining $m_w$ as the original audio track and $m_l$ as the audio degraded by one of the acoustic perturbations (detailed in Section~\ref{subsec:perturbations}).
\begin{equation*}
\begin{aligned}
\mathcal{L}_{\mathrm{CoPO}}
= -\log \sigma \Bigg(
&\beta \log
\frac{\pi_{\theta}(y_w \mid m_w,q)}
     {\pi_{\mathrm{ref}}(y_w \mid m_w,q)}
\\
&-
\beta \log
\frac{\pi_{\theta}(y_w \mid m_l,q)}
     {\pi_{\mathrm{ref}}(y_w \mid m_l,q)}
\Bigg)
\end{aligned}
\end{equation*}

\subsubsection{Anchored Preference Optimization Term}
A known pathology of the standard DPO formulation is that it optimizes solely for the relative reward margin; the model can spuriously increase this margin by reducing the likelihood of both the chosen and rejected responses, ultimately degrading the absolute generation quality of preferred outputs. To mitigate this likelihood degradation, the Anchored Preference Optimization term, $\mathcal{L}_{\mathrm{AncPO}}$, explicitly anchors the implicit reward of the chosen response to remain above a predefined threshold $\delta$~\cite{wang-etal-2024-mdpo}. 
\begin{equation}
\mathcal{L}_{\mathrm{AncPO}}
=
-\log \sigma \left(
\beta \log
\frac{\pi_{\theta}(y_w \mid m_{w},q)}
     {\pi_{\mathrm{ref}}(y_w \mid m_{w},q)}
-
\delta
\right)
\label{eq:ancpo}
\end{equation}
Consistent with the original mDPO implementation, we empirically set this anchor margin to $\delta = 0$ for all our experiments.

\subsubsection{Final objective}
Combining these components, the final mDPO objective function is defined as follows. 
\begin{equation}
\mathcal{L}_{\mathrm{mDPO}} = \mathcal{L}_{\mathrm{DPO}} + \mathcal{L}_{\mathrm{CoPO}} + \mathcal{L}_{\mathrm{AncPO}}
\end{equation}
As demonstrated in our ablation studies, each loss term plays a critical role in effectively mitigating multimodal hallucinations.

\subsection{Perturbations}
\label{subsec:perturbations}
In this work, we investigate five distinct acoustic perturbations, drawing inspiration from recent contrastive decoding literature that explores their efficacy in mitigating hallucinations~\cite{hsu2025reducing, grace2026adaptive}. These perturbations, some of which were implemented using the \textit{librosa} library~\cite{mcfee2015librosa}, are defined as follows:
\begin{itemize}
    \item \textbf{No Audio}: Masks the entire audio signal. This perturbation was shown to be highly effective for binary classification in Audio-Aware Decoding~\cite{hsu2025reducing} and subsequent studies~\cite{grace2026adaptive}.
    \item \textbf{Reverse}: Reverses the entire audio waveform temporally.
    \item \textbf{Random Noise}: Adds Gaussian noise to the audio at $\sigma_{wave}$ relative to waveform amplitude.
    \item \textbf{Time Masking}: Masks out intervals of time in audio, up to 15\% of the duration.
    \item \textbf{Frequency Masking}: Several random STFT(Short-Time Fourier Transform) bands are zeroed and reconstructed with ISTFT (Inverse Short-Time Fourier Transform).
\end{itemize}

\section{Related Work}
\label{sec:relatedwork}

\textbf{Audio QA and Advanced Fine-Tuning.} 
The recently concluded DCASE 2025 Challenge Task 5 highlighted the growing trend of leveraging large audio-language models (LALMs) for complex audio question answering. To achieve top performance, state-of-the-art systems rely heavily on diverse data augmentation strategies and advanced fine-tuning methodologies. For instance, recent approaches have utilized manual data curation~\cite{He_cuhk_t5_2025}, programmatic reformatting of annotations~\cite{Cai_ustc_2025}, and iterative pseudo-labeling~\cite{Chen_srcn_t5_2025}. These expanded datasets are then used to align models via Supervised Fine-Tuning (SFT) combined with Group Relative Policy Optimization (GRPO) or parameter-efficient Reinforcement Learning~\cite{Cai_ustc_2025, Chen_srcn_t5_2025, He_cuhk_t5_2025}. While these methods have proven highly effective for domain adaptation and alignment, we explore a different approach specifically designed to penalize multimodal hallucinations.

\textbf{Mitigating Multimodal Hallucinations.} 
To address modality-specific errors caused by language over-reliance, several training-free, inference-time techniques have been proposed~\cite{hsu2025reducing, grace2026adaptive, kuan2025can}. Contrastive decoding approaches like Audio-Aware Decoding~\cite{hsu2025reducing} dynamically adjust output logits by contrasting standard predictions against a negative branch (e.g., no-audio or blurred audio). While successful at penalizing language priors during generation, they do not correct the underlying acoustic representations. On the other hand, Preference Optimization methods like DPO~\cite{rafailov2023direct} directly update model weights, and recent extensions like Conditional Preference Optimization (mDPO)~\cite{wang-etal-2024-mdpo} explicitly target multimodal hallucinations in vision-language tasks by anchoring optimization to intact versus distorted images. 

In this work, we synthesize these paradigms. Rather than relying on inference-time corrections or standard RL, we adapt the mDPO framework to the audio domain. By utilizing the negative audio branches from contrastive decoding as the rejected inputs during preference learning, our approach internalizes these contrastive signals directly into the model's weights, providing a fundamental correction to acoustic grounding.

\section{Experimental Setup}
\label{sec:experimentalsetup}

\subsection{Datasets}
We evaluate our proposed methodology on two datasets. The first is the DCASE Challenge 2025 benchmark Task 5 (Audio Question Answering)~\cite{yang2026multidomain}. This multiple-choice dataset is divided into three domains: Bioacoustics QA, Temporal Soundscapes QA, and Complex QA. Because the Bioacoustics audio data is not currently publicly accessible, we restrict our evaluation to the Temporal and Complex QA subsets, yielding approximately 9,700 QA pairs. The dataset natively provides training and development splits; we further partition the development set equally to form our validation and testing sets. The second dataset is the AH Benchmark~\cite{kuan2025can}, which comprises three subcategories: AH Existence, AH Order, and AH Attribute. For our experiments, we isolate the AH Existence subset, consisting of 10,800 audio clips constructed by mixing three distinct audio events over a background track. Each clip is paired with multiple binary (yes/no) questions regarding the occurrence of specific events.

To apply the mDPO framework, we must construct a preference dataset consisting of paired text and paired audio. For every sample, we generate a tuple $(q, y_w, y_l, m_w, m_l)$, where a textual query $q$ is paired with a desired response $y_w$ and an incorrect/rejected response $y_l$. Simultaneously, the original clean audio $m_w$ is paired with its perturbed counterpart $m_l$. Thus, for every specific  perturbation evaluated, a completely new version of the training set is generated.

\subsection{Model}
Our primary experimental backbone is the Qwen2-Audio7B-Instruct~\cite{yang2024qwen2technicalreport}. For the DCASE 2025 dataset, the prompt is adapted to: ``Focus on the given audio and answer the following multiple-choice question. Respond with the letter of the correct answer (A, B, C, or D)''. To construct the negative responses required by the mDPO loss formulation in multiple-choice settings, we randomly sample one of the incorrect options. For the AH Existence dataset~\cite{kuan2025can}, we modify the instruction prompt to: ``Focus on the given audio and answer the following question with exactly one word: yes or no''.

We trained Qwen 2-7B-Instruct~\cite{yang2024qwen2technicalreport} on 5 epochs with a batch size of 4 (1 per GPU, with 4 GPUs). We additionally used LoRA\cite{hu2022lora} with a rank of 8 and an $\alpha$ of 16, where $\alpha$ is a scaling hyperparameter to control the weights of the model.

\subsection{Metrics}
The primary evaluation metric across all experiments is classification accuracy. For the DCASE 2025 dataset~\cite{yang2026multidomain}, we extract the predicted option letter and compare it against the correct label. In addition to overall accuracy, we independently report performance on the Temporal and Complex question subsets. For the AH Existence dataset~\cite{kuan2025can}, accuracy is computed by strictly matching the generated text against the ground-truth binary label.

\begin{table}[t]
\centering
\caption{Ablation on mDPO Loss terms with no audio perturbation on DCASE 2025 validation set.}
\label{tab:mdpo_results}

\begin{tabular}{lc}
\hline
Method & Overall Accuracy \\ \hline
Baseline & 56.0\% \\
mDPO & 71.4\% \\
$\quad - \mathcal{L}_{\mathrm{DPO}}$ & 70.4\% \\
$\quad - \mathcal{L}_{\mathrm{CoPO}}$ & 68.9\% \\
$\quad - \mathcal{L}_{\mathrm{AncPo}}$ & 64.0\% \\
\hline
\end{tabular}








\end{table}

\begin{table}[!t]
\centering
\caption{Beta values and corresponding accuracy with no audio perturbation on DCASE 2025 validation set.}
\label{tab:beta_values}
\begin{tabular}{cc}
\hline
\textbf{$\beta$} & \textbf{Overall Accuracy} \\
\hline

0.05 & 64.4\% \\
0.10 & 69.4\% \\
0.15 & 69.7\% \\
\textbf{0.20} & \textbf{71.4\%} \\
0.25 & 70.0\% \\

\hline
\end{tabular}

\end{table}

\section{Results}
\label{sec:results}

\begin{table}[t]
\centering
\caption{Time masking hyperparameter tuning on DCASE 2025 validation set.}
\label{tab:time_masking}
\begin{tabular}{lccc}
\hline
\shortstack{\textbf{Parameters}\\$(n_{\text{masks}},\; w_{\max})$} &
\shortstack{\textbf{Overall}\\\textbf{Accuracy}} &
\shortstack{\textbf{Temporal}\\\textbf{Accuracy}} &
\shortstack{\textbf{Complex}\\\textbf{Accuracy}} \\
\hline

Baseline &
56.0\% &
39.2\% &
62.3\% \\

(1,\ \ \ 0.1) &
69.8\% &
44.1\% &
79.3\% \\

(2,\ \ \ 0.1) & 
69.6\% &
42.8\% &
79.5\% \\

(2,\ \ \ 0.08) & 
69.8\% &
43.1\% &
\bf 79.6\% \\

(3,\ \ \ 0.1) & 
69.3\% &
42.2\% &
79.4\% \\

\bf (3,\ \ \ 0.08) & 
\textbf{70.4\%} &
\textbf{46.7\%} &
79.2\% \\

(3,\ \ \ 0.05) & 
68.2\% &
40.5\% &
78.4\% \\

\hline
\end{tabular}
\end{table}

\begin{table}[t]
\centering
\caption{Noise perturbation hyperparameter tuning on DCASE 2025 validation set.}
\label{tab:noise_results}
\begin{tabular}{lccc}
\hline
\textbf{$\sigma_{wave}$} &
\shortstack{\textbf{Overall}\\\textbf{Accuracy}} &
\shortstack{\textbf{Temporal}\\\textbf{Accuracy}} &
\shortstack{\textbf{Complex}\\\textbf{Accuracy}} \\
\hline

0 &
56.0\% &
39.2\% &
62.3\% \\

0.01 &
69.0\% &
42.5\% &
78.8\% \\

\textbf{0.02} &
\textbf{69.5\%} &
43.1\% &
\textbf{79.3\%} \\

0.03 &
69.2\% &
\textbf{43.8\%} &
78.6\% \\

0.05 &
68.3\% &
\textbf{43.8\%} &
77.4\% \\

\hline
\end{tabular}
\end{table}

\begin{table}[t]
\centering
\caption{Qwen2 accuracy on DCASE 2025 {test set} across different perturbations.
}
\label{tab:qwen2_perturbations}
\begin{tabular}{@{}l@{\hspace{0.02\linewidth}}c@{\hspace{0.02\linewidth}}c@{\hspace{0.02\linewidth}}c@{}}
\hline
\textbf{Perturbation} &
\shortstack{\textbf{Overall}\\\textbf{Accuracy}} &
\shortstack{\textbf{Temporal}\\\textbf{Accuracy}} &
\shortstack{\textbf{Complex}\\\textbf{Accuracy}} \\
\hline
Baseline &
56.0\% &
39.2\% &
62.3\% \\
No Audio &
66.7\% &
43.1\% &
75.5\% \\
Reverse &
{67.1\%} &
\textbf{44.1\%} &
75.6\% \\
Random Noise &
{67.4\%} &
43.8\% &
76.2\% \\
Time Masking &
66.6\% &
39.5\% &
76.7\% \\
Frequency Masking &
{67.3\%} &
43.4\% &
76.2\% \\
Triple Mixed Perturbations &
\textbf{70.0\%} &
43.8\% &
\bf 80.0\% \\
\hline
DCASE Team 1 Results~\cite{He_cuhk_t5_2025} &
&
60.4\% &
83.3\% \\
\hline
DCASE Team 2 Results~\cite{Cai_ustc_2025} &
&
62.6\% &
80.1\% \\
\hline
\end{tabular}
\end{table}

\begin{table}[t]
\centering
\caption{Qwen2 accuracy on AH Existence across different perturbations.}
\label{tab:qwen2_ah_existence_perturbations}
\begin{tabular}{lc}
\hline
\textbf{Perturbation} &
\textbf{Test Accuracy} \\
\hline

Baseline
& 55.6\% \\

No Audio
& 81.3\% \\

Reverse
& \bf 83.0\% \\

Random Noise
& 81.2\% \\

Frequency Masking
& 79.2\% \\
\hline
Contrastive Decoding~\cite{grace2026adaptive} & 76.7\% \\
\hline
\end{tabular}
\end{table}

\subsection{Ablation Studies}
\label{ssec:ablationstudies}











\subsubsection{Effect of loss terms in mDPO}
\label{sssec:removeterms}
We initially performed an ablation study to quantify the individual contribution of each component within the mDPO objective (Table \ref{tab:mdpo_results}). This experiment utilized the no-audio perturbation and held $\beta$ constant at 0.20. The results demonstrate that the three loss terms are highly complementary to each other; applying the full mDPO objective significantly improved performance, increasing accuracy from a baseline of 56.0\% to 71.4\%, whereas removing any single term degraded performance. Most notably, removing the Anchor loss ($\mathcal{L}_{\mathrm{AncPo}}$) caused the most substantial degradation, dropping accuracy to 64.0\%. This highlights its critical role in preventing likelihood degradation and preserving the absolute generation quality of preferred outputs.

\subsubsection{Testing Beta Values}
\label{sssec:betavalues}
We subsequently conducted an ablation study to determine the optimal $\beta$ parameter for the mDPO objective. As demonstrated in Table \ref{tab:beta_values}, setting $\beta = 0.20$ yielded the highest accuracy; consequently, we adopted this value for all subsequent experiments.

\subsubsection{Perturbation Ablation Studies}
We conducted further ablation studies to determine the optimal hyperparameters for the time masking and random noise perturbations. As detailed in Table \ref{tab:time_masking}, configuring time masking with $n_{\text{masks}} = 3$ and $w_{\max} = 0.08$ achieved the highest accuracy. Regarding the noise perturbation (Table \ref{tab:noise_results}), a noise level of $\sigma_{wave} = 0.02$ maximized overall performance, while slightly higher noise levels ($\sigma_{wave} \in \{0.03, 0.05\}$) maximized temporal accuracy at 43.8\%.

\subsection{Test Results}

On the DCASE 2025 dataset~\cite{yang2026multidomain}, we independently evaluated each perturbation while training Qwen2~\cite{yang2024qwen2technicalreport}, yielding the test set accuracies presented in Table~\ref{tab:qwen2_perturbations}. The reverse, random noise, and frequency masking perturbations individually achieved the highest performance. To leverage these findings, we constructed a combined training set---termed \textit{Triple Mixed Perturbations}---by tripling each observation so that every audio sample is paired with each of these three top-performing perturbations. As shown in Table~\ref{tab:qwen2_perturbations}, training on this combined dataset yielded the highest overall accuracy of 70.0\%.

Furthermore, the bottom section of Table~\ref{tab:qwen2_perturbations} presents the performance of two top-performing teams from the DCASE 2025 Challenge. Although these teams utilized external data for finetuning their LALMs, our perturbation-based approach achieves comparable results, particularly on the Complex task.

Additionally, we finetuned Qwen2 on the AH Existence dataset across various perturbations, directly transferring the optimal hyperparameters discovered on DCASE without further tuning. As reported in Table \ref{tab:qwen2_ah_existence_perturbations}, the reverse perturbation achieved the highest accuracy at 83.0\%. Notably, our preference optimization approach significantly outperforms inference-time contrastive decoding~\cite{grace2026adaptive}, which achieved 76.7\% using the Qwen2 backbone.

\section{Conclusions}
In this work, we explored extending the mDPO framework~\cite{wang-etal-2024-mdpo} to Large Audio Language Models (LALMs) to mitigate hallucinations and improve multimodal grounding. By applying acoustic perturbations during preference optimization, we demonstrated significant performance gains across two benchmarks. On the DCASE 2025 Challenge dataset, combining three distinct perturbations (Triple Mixed) improved overall accuracy from 56.0\% to 70.0\%, while the reverse perturbation specifically boosted temporal accuracy to 44.1\% from 39.2\%. Similarly, on the AH Existence dataset~\cite{kuan2025can}, the reverse perturbation elevated accuracy from 55.6\% to 83.0\%. These findings underscore the effectiveness of incorporating acoustic perturbations directly into the finetuning process, offering a more robust alternative to inference-time contrastive decoding. For future work, we plan to extend this methodology to open-ended generative tasks, such as audio captioning, and explore adaptive perturbation strategies that dynamically adjust distortion difficulty during training.

\cleardoublepage

\bibliographystyle{IEEEbib}
\bibliography{strings}

\end{document}